%% file: main.tex
\expandafter\def\csname ver@fixltx2e.sty\endcsname{}

\documentclass[a4paper,oneside,preprint,12pt]{elsarticle}

\input{Config/packages}
\input{Config/settings}
\input{Config/macros}

\begin{document}

\begin{frontmatter}

    \title{Synchrotron-based Photonuclear Neutron Source for Energy, Medicine and Radiation Testing}

    \author[Poli,Bicocca,Khalifa1,Khalifa2]{Antonio Cammi\,\orcidlink{0000-0003-1508-5935}\,$^{*,}$}
    \author[Poli,Bicocca]{Lorenzo Loi\,\orcidlink{0009-0003-7904-3731}\,}
    \author[Poli]{Andrea Missaglia\,\orcidlink{0009-0005-9552-4847}\,}
    \author[Frascati]{David Alesini\,\orcidlink{0000-0002-6778-882X}\,}
    \author[PSI]{Hans Heinrich Braun\,\orcidlink{0009-0006-4356-4334}\,}

    \cortext[cor1]{Corresponding author. Email address: antonio.cammi@polimi.it}
    
    \address[Poli]{Politecnico di Milano, Department of Energy, Via La Masa 34, Milano 20133, Italy}
    \address[Bicocca]{INFN, Sezione di Milano Bicocca, 20126 Milan, Italy}
    \address[Khalifa1]{Department of Mechanical and Nuclear Engineering, Khalifa University, Abu Dhabi, 127788, United Arab Emirates}
    \address[Khalifa2]{Emirates Nuclear Technology Centre (ENTC), Khalifa University of Science and Technology, Abu Dhabi, 127788, United Arab Emirates}
    \address[Frascati]{INFN–Laboratori Nazionali di Frascati, Via E. Fermi 54, Frascati (Roma) 00044, Italy}
    \address[PSI]{Paul Scherrer Institut, Forschungsstrasse 111 5232 Villigen PSI, Switzerland}

\begin{abstract}
Neutron source availability is constrained by the declining number of research reactors and the challenging cost of spallation facilities, while compact accelerator-based sources remain limited to beam powers below 100~kW due to severe thermal loads from charged particle-target interactions. Here we demonstrate that synchrotron-driven photonuclear production overcomes this fundamental limitation through architectural decoupling: electrons circulating in a storage ring generate MeV-range synchrotron radiation that drives external neutron production targets, completely isolating targets from primary charged-particle power. This separation enables continuous-wave photon beam powers exceeding 200~kW per beamline while drastically reducing localized heating and reducing thermal shocks. Through systematic Monte Carlo optimization across three independent codes (OpenMC, MCNPX, FLUKA), we identify optimal target configurations delivering single-beamline neutron intensities of $3$--$13 \times 10^{14}$~n/s with inherent spectral tunability: low-Z targets (Be, D$_2$O) produce epithermal-to-thermal spectra for medical applications, while high-Z converters (W, U) generate fast neutrons for subcritical reactors. Multi-beamline architecture enables facility-scale outputs exceeding $6 \times 10^{16}$~n/s, being competitive with research reactors, while supporting simultaneous parallel operation of application-specific stations.
\end{abstract}


\end{frontmatter}


\input{Sections/01_introduction}
\input{Sections/03_results}
\input{Sections/04_discussion}

\input{Sections/05_conclusions}
\input{Sections/02_methodology}

\input{Sections/xx_symbols}

\bibliography{bibliography.bib}


\end{document}

%% file: Config/packages.tex
\usepackage[utf8]{inputenc}
\usepackage{microtype} 
\usepackage{lmodern}

\usepackage{amsmath}
\usepackage{amsthm}
\usepackage{amssymb}
\usepackage{amsfonts}
\usepackage{mathtools}
\usepackage{bm}
\usepackage{siunitx}

\usepackage{booktabs} 
\usepackage{tabularx}
\usepackage{ragged2e}
\usepackage{longtable}
\usepackage{multirow}
\usepackage{multicol}
\usepackage{rotating}

\usepackage[dvipsnames]{xcolor}
\usepackage{graphicx}
\usepackage{epstopdf}
\usepackage{subcaption}
\usepackage{tikz}
\usepackage{tikz-3dplot}
\usetikzlibrary{fit, shapes, arrows, backgrounds, calc, arrows.meta, positioning, decorations.pathreplacing, fadings}

\usepackage[ruled,vlined]{algorithm2e}

\usepackage{url}
\usepackage{flushend} 
\usepackage{parskip}
\usepackage{framed}
\usepackage{emptypage}
\usepackage[printonlyused]{acronym}
\usepackage{comment}
\usepackage{lineno} 

\usepackage[hyperfootnotes=true, colorlinks=true, linkcolor=blue, bookmarks=true, citecolor={red}, urlcolor=cyan]{hyperref}
\usepackage{orcidlink}

%% file: Config/settings.tex
\usepackage{geometry}
\journal{}

%% file: Config/macros.tex
\definecolor{bluePolimi}{RGB}{22, 44, 80}
\definecolor{lightBluePolimi}{RGB}{91, 122, 172}
\definecolor{greenPolimi}{RGB}{0, 110, 0}
\definecolor{redPolimi}{RGB}{180, 0, 0}

\newcolumntype{M}[1]{>{\centering\arraybackslash}m{#1}}

\tikzset{
    beamColor/.style={red!80!black},
    neutronColor/.style={blue!60!cyan},
    targetBody/.style={top color=gray!10, bottom color=gray!40, middle color=white, shading angle=0},
    targetFace/.style={fill=gray!20, draw=gray!60},
    controlSurf/.style={dashed, color=gray!40, thin},
    myLabel/.style={font=\sffamily\small, text=black!80}
}

%% file: Sections/01_introduction.tex
\section{Introduction}\label{sec:intro}

Neutron sources play a critical role in diverse scientific and technological domains, from fundamental neutron scattering studies and materials characterization to medical isotope production, boron neutron capture therapy (BNCT) and accelerator-driven subcritical reactor (ADS) research~\cite{Kiyanagi2021}. While nuclear reactors remain the dominant high-flux solution, their declining global availability, regulatory complexity, and operational cost have driven sustained interest in accelerator-based alternatives~\cite{Bauer2001, Rush_2015}. The primary option lies in large-scale spallation facilities, able to deliver neutron source rates exceeding $10^{17}$~n/s but requiring substantial capital investment and complex infrastructure. 

\begin{figure}[t!]
    \centering
    \includegraphics[width=0.8\linewidth]{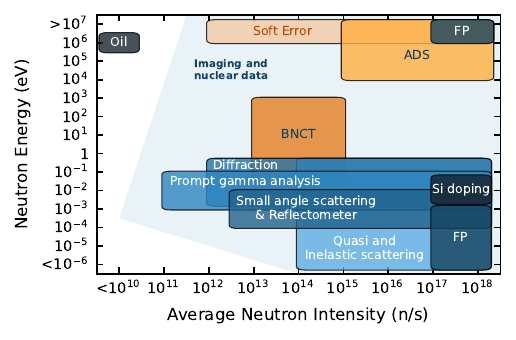} 
    \caption{Classification of neutron sources and their application domains based on neutron intensity. Data adapted from~\cite{Kiyanagi2021}. FP = Fundamental Physics; BNCT = Boron Neutron Capture Therapy; ADS = Accelerator Driven Systems.}
    \label{fig:kiyanagi_gap}
\end{figure}

The mismatch between the expanding application diversity (see Figure~\ref{fig:kiyanagi_gap}) and the constrained source availability, demands neutron sources that are not only intense, but also flexible and accessible. Compact Accelerator Neutron Sources (CANS) have emerged as a promising solution, explicitly designed to bridge this gap with target intensities of $10^{12}$--$10^{14}$~n/s~\cite{Carpenter2019, Anderson2016}. CANS rely on two primary approaches: proton-driven ($p,n$) reactions and electron-driven photonuclear ($\gamma,xn$) reactions~\cite{Findlay_1990}.

Despite decades of development, both proton- and electron-driven CANS face a fundamental common constraint: achievable beam powers are limited to approximately 100~kW, restricting their scalability and accessible neutron intensity regimes~\cite{Gutberlet2025, Gohar_OSTI, Herrador2025}. This limitation arises from the physics of charged particle-material interactions, where beam energy is deposited through ionization processes, creating localized pulsed heat loads that necessitate complex thermal management solutions. The severity and specific manifestations of these constraints, however, differ markedly between the two conventional designs.\\

Proton-driven CANS, using low-energy protons (E$_p$ $<$ 100~MeV) on lithium or beryllium targets, represent the most mature compact technology with favorable $(p,n)$ reaction yields. Protons lose energy volumetrically via ionization, creating intense peak energy deposition densities at the Bragg peak that generate severe thermo-mechanical stresses. Managing the resulting localized heating requires complex engineering solutions such as rotating targets or liquid-metal cooling loops. Target longevity is further compromised by material degradation: accumulation of reaction by-products (hydrogen, helium) leads to blistering and radiation-induced swelling, imposing strict limits on target lifetime~\cite{Lai2022}. Operational stability is also challenged by frequent beam trips that prevent delivery of the continuous flux required for many applications~\cite{Anderson2016}.

Electron-driven CANS, while offering operational advantages such as the widespread availability of electron LINACs and favorable economics for neutron rates below $10^{16}$~n/s when compared to spallation facilities~\cite{Ridikas_2001}, face even more severe beam power restrictions. This technology follows a two-step approach for neutron generation: first, electrons generate gamma rays on high-$Z$ converters (tungsten, tantalum) through bremsstrahlung; second, these photons induce $(\gamma,xn)$ reactions on either low-$Z$ materials (deuterium, beryllium) or high-$Z$ targets (tungsten, tantalum, uranium). The conversion process is intrinsically inefficient, with up to half of the electron beam power deposited as heat in the converter itself~\cite{Miloshevskii1998}, leaving only the remaining fraction as photon beam power available for neutron production. Consequently, operational electron-driven CANS are restricted to beam powers below 10~kW with neutron intensities in the range $1\times 10^{11}$--$1\times10^{13}$~n/s~\cite{Herrador2025}. Even high-power systems such as KIPT, operating near 100~kW electron beam power, achieve only $\sim 3 \times 10^{14}$~n/s~\cite{Gohar_OSTI, Mytsykov2019}. This leaves the upper medium-intensity regime ($10^{14}$--$10^{16}$~n/s) largely inaccessible to conventional compact architectures.

\begin{figure}[t!]
    \centering
    \includegraphics[width=0.9\linewidth]{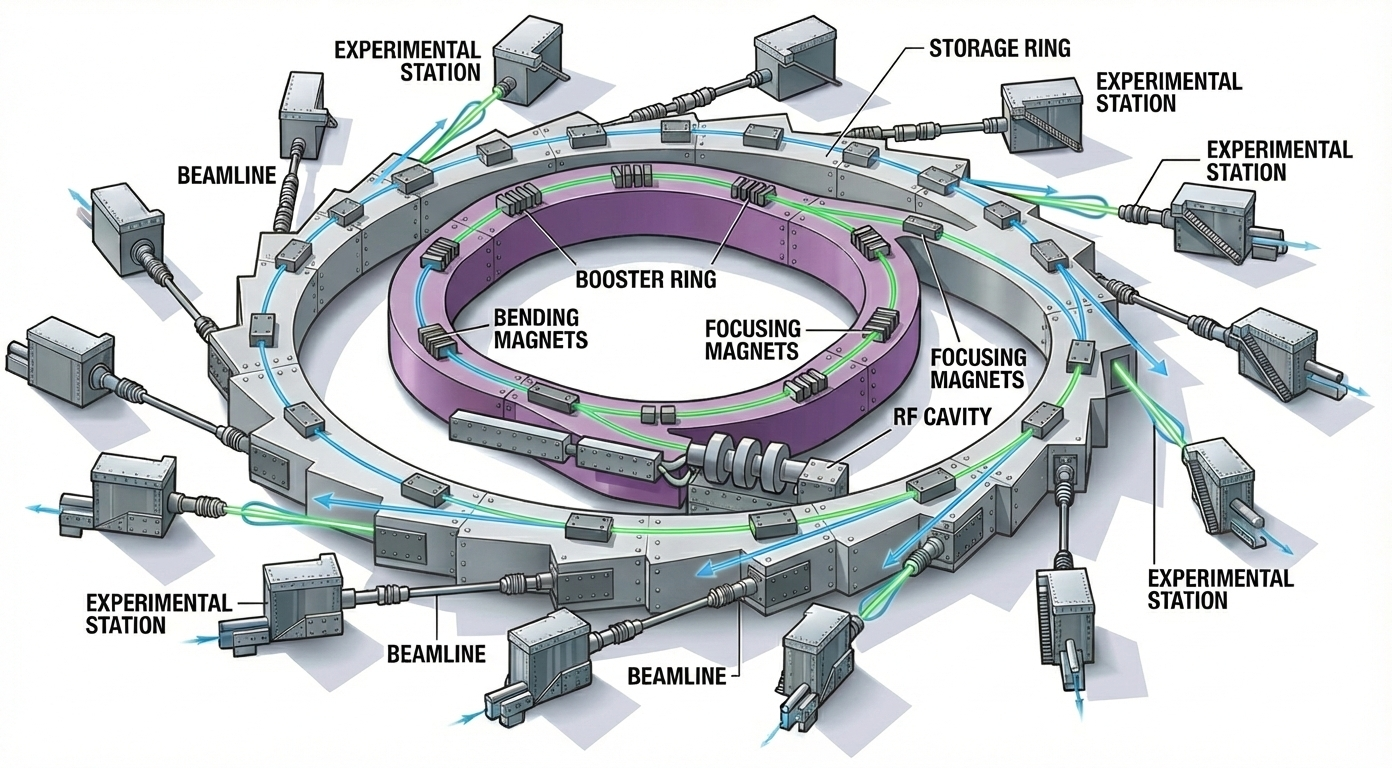}
    \caption{Illustrative example of a multi-beamline synchrotron-based facility, where a single storage ring feeds multiple photon beamlines.}
    \label{fig:synchrotron_concept}
\end{figure}

A third, less explored, pathway involves synchrotron-driven neutron generation. Early feasibility studies at Japan's SPring-8 facility evaluated photoneutron yields from beryllium targets~\cite{SPRING8, GRYAZNYKH_2000, Asano2005}, but the synchrotron radiation spectrum remained predominantly below the beryllium production threshold (critical energy $\sim$0.42~MeV), limiting neutron generation efficiency. By substantially increasing the synchrotron photon energy to a 10~MeV critical energy, we re-evaluate synchrotron capabilities as competitive neutron sources~\cite{alesini_workshop_INFN_E_A}.\\

We introduce SYNERGY (SYnchrotron-driven NEutron source for Research, energy Generation and therapY), a novel photonuclear neutron production concept that overcomes traditional beam power limitations through architectural decoupling of charged-particle acceleration from neutron production. The concept exploits a storage ring confining high-energy electrons in a closed orbit, where bending magnets continuously generate synchrotron radiation. The emitted photon beam is extracted tangentially and transported to an external neutron-production target (Figure~\ref{fig:synchrotron_concept}). Because radiated energy must be continuously replenished to maintain the stored beam, the maximum extractable photon power is bounded by the energy restoration capacity of the radio-frequency (RF) cavities. Considering realistic RF-to-beam power transfer efficiencies achievable in modern storage rings, this topological separation enables photon beam powers exceeding 200~kW per beamline~\cite{FCC_report, l_loi_2026_20804002}.

This architectural decoupling has profound implications for thermal management. Unlike conventional CANS where charged particles directly interact with neutron-production targets, the synchrotron design isolates the target from primary charged-particle power, exposing it exclusively to the photon beam. MeV-range photons exhibit significantly longer interaction lengths and deposit substantially less energy per interaction than charged particles, reducing localized heat loads and distributing energy deposition more uniformly throughout the target volume~\cite{Byun2021}. Also, the continuous-wave (CW) nature of synchrotron radiation, contrasting sharply with pulsed operation, further eliminates thermal shocks and lowers mechanical fatigue. This combination fundamentally relaxes the thermal bottleneck constraining both proton- and electron-driven CANS without requiring exotic engineering solutions.

Also, the synchrotron architecture inherently supports multi-user operation. A sufficiently large storage ring simultaneously feeds numerous independent photon beamlines distributed around the accelerator circumference. Multiple photoneutron stations may operate in parallel, each optimized for specific neutron spectra and applications, such as fast neutron fields for radiation testing and subcritical reactor coupling, epithermal spectra for BNCT, and moderated fields for neutron scattering. By distributing neutron production across this multi-beamline architecture, the system achieves total neutron source rates competitive with research reactors while maintaining manageable per-target thermal loads. This capability addresses the broad application requirements illustrated in Figure~\ref{fig:kiyanagi_gap}, dramatically improving scientific reach and cost-effectiveness relative to single-beam drivers.\\

This work presents the first comprehensive feasibility assessment of SYNERGY as a neutron production technology. Through systematic parametric Monte Carlo analysis validated across three independent codes (OpenMC, MCNPX, FLUKA), we perform geometry optimization and quantify neutron source rates and energy spectra achievable from uranium, tungsten, beryllium, and heavy water targets under realistic synchrotron operational parameters. We demonstrate single-beamline neutron intensities from $2.8 \times 10^{14}$ to $1.3 \times 10^{15}$~n/s, with spectra naturally tailored to distinct applications through target material selection. When extended to a 50-beamline configuration representative of state-of-the-art facilities, the total facility intensity exceeds $6.0 \times 10^{16}$~n/s, positioning synchrotron-driven sources as a credible alternative in the medium-to-high flux regime. We assess these results in the context of existing neutron source capabilities and application requirements, demonstrating the potential for simultaneous support of accelerator-driven subcritical systems, medical isotope production, radiation hardness testing, and BNCT within a single facility.

%% file: Sections/03_results.tex
\section{Results}\label{sec:results}

We now present the key findings of the performance assessment and the parametric optimization of the SYNERGY concept, organized in three parts: target geometry optimization, spectral characteristics with verification across Monte Carlo codes, and estimation of absolute neutron rates. The last results are derived assuming a 200~kW photon beam power per beamline, representative of achievable synchrotron radiation outputs.

\subsection{Target Geometry Optimization}

\begin{figure*}[h!]
\centering
\includegraphics[width=1\linewidth]{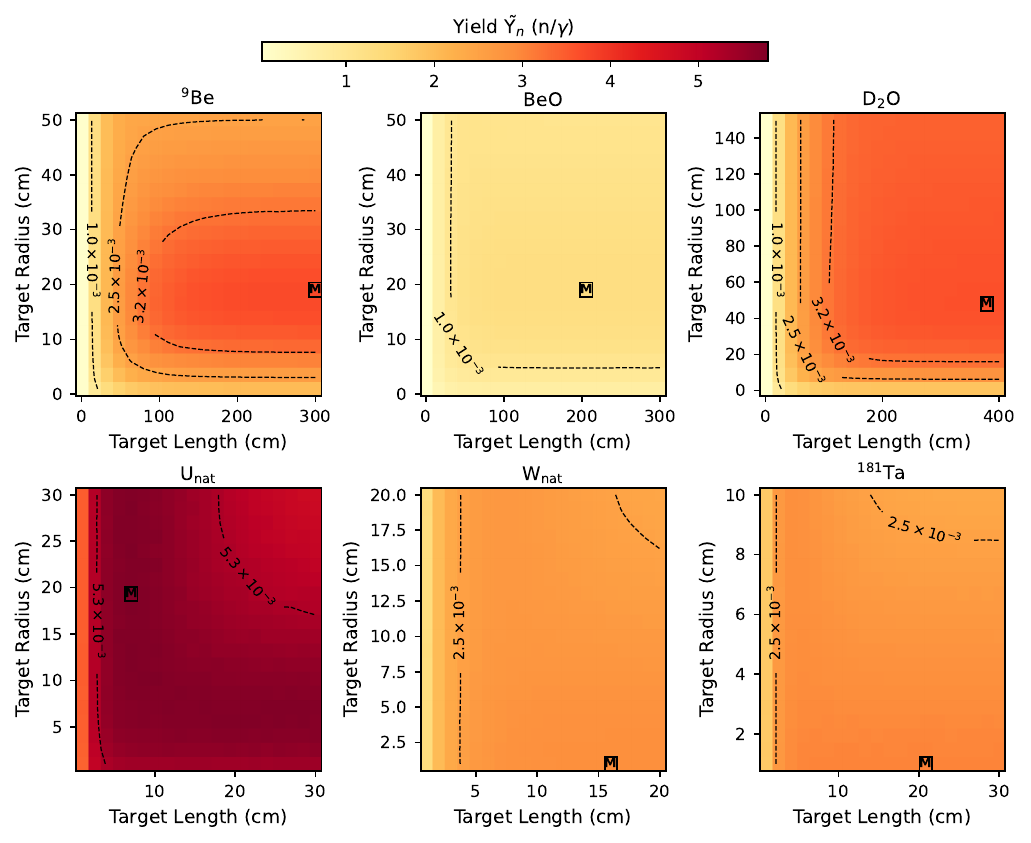}
\caption{Photoneutron-yield scan over target radius and length for all materials. The color scale indicates the magnitude of $\Tilde{Y}_n$ (neutrons/source photon), and the marker \textbf{M} indicates the optimal geometric configuration.}
\label{fig:Yn_total}
\end{figure*}

Target geometry critically influences neutron production efficiency. For each candidate material, we performed a systematic parametric sweep over target radius ($R$) and length ($L$), evaluating the photoneutron yield ($\Tilde{Y}_n$, defined as neutrons produced per source photon within the target) as a function of these parameters. Figure~\ref{fig:Yn_total} illustrates representative optimization surfaces for beryllium, beryllium oxyde, heavy water, tungsten, tantalum and natural uranium targets, where marker \textbf{M} indicates the configuration maximizing total photoneutron generation.

The analysis reveals distinct physical regimes governing optimal dimensions, driven by the competition between photon attenuation and neutron transport. High-Z materials (natural uranium, tungsten, tantalum) exhibit rapid yield saturation due to their high density and large photoatomic cross-sections, which fully attenuate the incident photon beam within a relatively short penetration depth. Tungsten and tantalum exploit the $(\gamma, n)$ reaction channel, while natural uranium additionally unlocks $(\gamma, 2n)$ and photofission, enhancing the source capabilities. Consequently, optimal dimensions are compact—lengths of 10--25~cm. A further volume increase would introduce detrimental neutron self-shielding rather than enhanced production.

In contrast, low-Z materials (heavy water, beryllium, beryllium oxide) require significantly larger volumes to approach saturation, with optima exceeding 2~m in length and 50~cm in radius. This reflects the longer photon mean free path (dominated by Compton scattering) and the need to minimize neutron leakage through extensive in-target moderation and thermalization. These materials naturally generate epithermal-to-thermal neutron spectra valuable for applications such as neutron scattering and BNCT.

Table~\ref{tab:result_optimization} summarizes the identified optimal dimensions ($R^{\mathrm{opt}}$, $L^{\mathrm{opt}}$) and corresponding maximum yields for all candidate materials. These geometric configurations serve as the reference design for subsequent spectral analysis. For more details about the source description, refer to Section~\ref{sec:synchrotron_methods}.

\begin{table}[ht]
\centering
\caption{Total photoneutron yields ($\Tilde{Y}_n^{\mathrm{opt}}$) and the corresponding geometric parameters (radius $R^{\mathrm{opt}}$ and length $L^{\mathrm{opt}}$) resulting from the parametric optimization sweep. Values were calculated using OpenMC with $10^7$ primary photon histories.}
\label{tab:result_optimization}
\begin{tabular}{@{}lccc@{}}
\toprule
Target & $\Tilde{Y}_n^{\text{opt}}$ ($10^{-3}$ n/$\gamma$) & $R^{\text{opt}}$ (cm) & $L^{\text{opt}}$ (cm) \\
\midrule
$^9$Be           & $3.703 \pm 0.002$ & 19.1 & 300.0 \\
BeO              & $1.297 \pm 0.001$ & 19.1 & 205.6 \\
D$_2$O           & $3.618 \pm 0.001$ & 48.1 & 379.0 \\
U$_{\text{nat}}$ & $5.792 \pm 0.011$ & 19.3 & 7.1   \\
W$_{\text{nat}}$ & $2.865 \pm 0.003$ & 1.0  & 16.0   \\
$^{181}$Ta       & $3.029 \pm 0.002$ & 1.0  & 20.8   \\
\bottomrule
\end{tabular}
\end{table}

\subsection{Spectral Analysis and Cross-Code Validation}

\begin{figure*}[h!]
\centering
\includegraphics[width=1\linewidth]{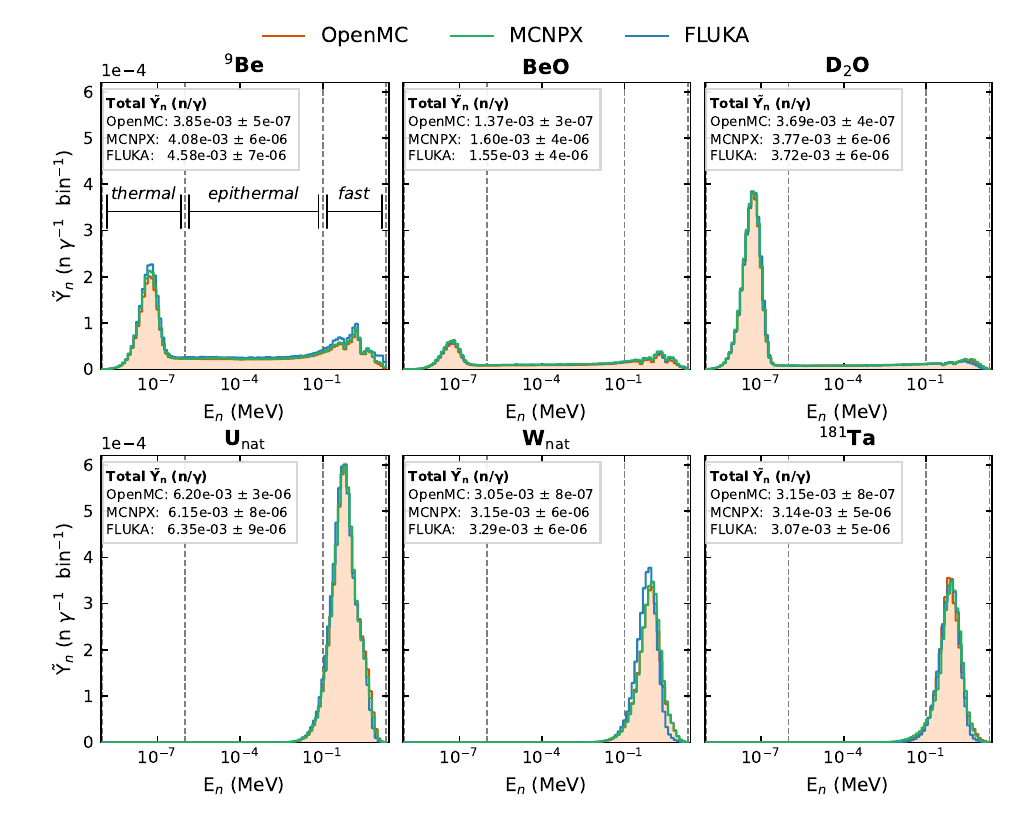}
\caption{Comparison of neutron energy distribution for the optimal target geometries using OpenMC, MCNPX, and FLUKA. The total photoneutron yield is reported in the dedicated chart for each code.}
\label{fig:Yn_best_E}
\end{figure*}

To verify the robustness of our results, we performed parallel simulations using three independent Monte Carlo codes: OpenMC (with photonuclear capabilities\cite{SteinRepo}), MCNPX 2.7, and FLUKA 4-5. Figure~\ref{fig:Yn_best_E} presents the neutron energy distributions calculated by all three codes for each candidate target material in its optimal configuration.

The codes demonstrate excellent agreement in both spectral shape and total yield. Relative deviations in integrated neutron production remain within $\pm 10\%$ for all materials except beryllium, where OpenMC and FLUKA differ by $\sim$20\%. Spectral shapes are consistent across codes, with statistical uncertainties maintained below 0.2\% through $10^8$ photon histories per simulation. This multi-code convergence confirms that the reported performance metrics are not artifacts of particular nuclear data libraries or physics implementations.

Importantly, the benchmarked spectra are naturally divided into two distinct categories aligned with different application requirements. High-Z targets (U, W, Ta) emit predominantly fast neutrons peaked between 0.1~MeV--1~MeV, characteristic of evaporation neutrons from $(\gamma, xn)$ and photofission reactions with minimal in-target moderation. These hard spectra are well-suited for radiation damage studies, accelerator-driven systems, and fast neutron imaging. In contrast, low-Z targets (D$_2$O, Be, BeO) exhibit pronounced thermal and epithermal components in the emerging spectrum. It is important to stress that the underlying photonuclear production mechanism is identical across all target materials: neutrons are generated with a fast evaporation spectrum regardless of the converter's atomic number, as detailed in \cite{loi2026photonuclearneutronproductionopenmc}. The softening observed for D$_2$O, Be, and BeO does not originate from a different emission process, but rather from the optimal target thickness required to maximize neutron yield. For these low-$Z$ moderating materials, the yield-optimized geometry extends over several neutron mean free paths, so that fast neutrons produced throughout the target volume undergo substantial elastic scattering and slowing-down before escaping. The exiting spectrum is therefore thermalized within the target itself, with the target simultaneously acting as neutron converter and moderator. High-$Z$ targets, by contrast, are optimized at thicknesses comparable to the photon attenuation length, insufficient to appreciably moderate the fast neutrons before they escape, so the hard evaporation spectrum is preserved. This distinction demonstrates how target material and geometry selection provides inherent spectral tunability without the need for additional, external moderation infrastructure.

For the remainder of this work, we adopt OpenMC results as the reference baseline, as they consistently yield the lowest integrated neutron production rates among the three codes. This choice ensures conservative performance estimates, with the understanding that alternative code outcomes may predict 5--10\% higher production rates.

\subsection{Absolute Production Rates and SYNERGY Performances}
To assess realistic performances, the calculated yields were normalized to a nominal synchrotron beam power of 200~kW per beamline, corresponding to an incident photon intensity of $2.02\times10^{17}$~$\gamma$/s in the 1--37~MeV range. Table~\ref{tab:photoneutron_flux_by_range} presents the absolute neutron source rates for each optimized target configuration, decomposed into three application-relevant energy groups: thermal flux (1~meV--1~eV) for neutron scattering and capture therapies; epithermal flux (1~eV--0.1~MeV) critical for clinical BNCT requirements; and fast flux (0.1--20~MeV) essential for subcritical reactor systems and radiation damage studies.

\begin{table*}[t]
\centering
\caption{Absolute neutron source rates ($S_n$) for optimized target materials, normalized to a primary synchrotron photon intensity of $S_\gamma = 2.02 \times 10^{17}$~$\gamma$/s and with details given in Section~\ref{sec:synchrotron_methods}. Data are categorized into thermal, epithermal, and fast energy ranges based on high-statistics OpenMC simulations ($10^8$ histories). Statistical uncertainties are below 0.2\% for all entries.}
\label{tab:photoneutron_flux_by_range}
\begin{tabular}{@{}lcccc@{}}
\toprule
& \multicolumn{4}{c}{Neutron Source Rate $S_n$ ($10^{14}$ n/s)} \\
\cmidrule(lr){2-5}
Material & Thermal & Epithermal & Fast & Total \\
& [1 meV -- 1 eV] & [1 eV -- 0.1 MeV] & [0.1 MeV -- 20 MeV] & [1 meV -- 20 MeV] \\
\midrule
$^9$Be           & 3.6    & 2.3    & 1.9    & 7.8  \\
BeO              & 1.0    & 0.97   & 0.76   & 2.8  \\
D$_2$O           & 6.0    & 0.88   & 0.56   & 7.4  \\
U$_{\text{nat}}$ & $<10^{-7}$ & 0.63   & 12.0   & 13.0 \\
W$_{\text{nat}}$ & $<10^{-7}$ & 0.21   & 6.0    & 6.1  \\
$^{181}$Ta       & $<10^{-7}$ & 0.23   & 6.1    & 6.4  \\
\bottomrule
\end{tabular}
\end{table*}

The results reveal a clear trade-off between total intensity and spectral characteristics. Natural uranium targets deliver the highest integrated source strength of $1.3 \times 10^{15}$~n/s per beamline, dominated by fast neutrons ($1.2 \times 10^{15}$~n/s, $\sim$90\% of total yield) from photofission and $(\gamma, xn)$ reactions, with negligible thermal component. High-Z non-fissile converters (W, Ta) achieve comparable performance at $6.1$--$6.4 \times 10^{14}$~n/s, similarly producing hard spectra with $>$95\% fast neutrons. In contrast, low-Z moderating targets exhibit fundamentally different spectral characteristics: beryllium produces $7.8 \times 10^{14}$~n/s with substantial thermal ($3.6 \times 10^{14}$~n/s, 46\%) and epithermal ($2.3 \times 10^{14}$~n/s, 29\%) components; heavy water generates $7.4 \times 10^{14}$~n/s dominated by thermal neutrons ($6.0 \times 10^{14}$~n/s, 81\%); and beryllium oxide yields $2.8 \times 10^{14}$~n/s with balanced thermal-epithermal-fast contributions. Critically, these soft-spectrum targets deliver their thermal and epithermal fluxes without external moderation infrastructure, providing application-ready neutron fields directly at the production point. This spectral diversity enables target material selection to be directly matched to application requirements, as detailed in Section~\ref{sec:discussion}.\\

Extending these single-beamline results to a full-scale synchrotron facility reveals the system-level capabilities of the SYNERGY concept. Representative large synchrotron facilities operate 40--60 independent beamlines simultaneously, each extracting photon power from dedicated insertion devices or bending magnets. Assuming a configuration of 50 operational beamlines, each delivering 200~kW photon power to an optimized neutron production target, the achievable total facility source rate spans:

\begin{itemize}
    \item \textbf{Moderating target (D$_2$O):} $S_n^{\text{total}} \approx 3.7 \times 10^{16}$~n/s (thermal: $3.0 \times 10^{16}$~n/s)
    \item \textbf{Fissile target (U$_{\text{nat}}$):} $S_n^{\text{total}} \approx 6.5 \times 10^{16}$~n/s (fast: $6.0 \times 10^{16}$~n/s)
\end{itemize}

These results position SYNERGY in the medium-to-high intensity regime, competitive with research reactors while offering superior spectral flexibility through parallel operation of diverse target stations. The implications of these source rates for specific applications are examined in Section~\ref{sec:discussion}.

%% file: Sections/04_discussion.tex
\section{Discussion}
\label{sec:discussion}

The results presented in Section~\ref{sec:results} demonstrate that synchrotron-driven photonuclear neutron production can overcome the fundamental beam power limitations that have constrained compact neutron sources for decades. By decoupling charged-particle acceleration from neutron generation, the SYNERGY concept achieves single-beamline intensities of $3$--$13 \times 10^{14}$~n/s and facility-scale outputs exceeding $6 \times 10^{16}$~n/s. These performance levels, previously accessible only through large spallation facilities or research reactors, are now delivered with inherent multi-user capability and spectral tunability. In this section, we aim to highlight the \textit{flexibility} of SYNERGY, comparing the single beam-line performance against the requirements of key application domains, understanding the position of this innovative facility in the broader landscape of neutron source technologies. Table~\ref{tab:synergy_facility_comparison} provides a comprehensive comparison with existing facilities, while the following discussion examines implications for subcritical reactor systems, medical applications, and radiation damage studies. We conclude by identifying critical technical challenges and outlining directions for experimental validation and engineering development.

\paragraph{Subcritical Reactors and Transmutation}
ADS require sustained external neutron sources to maintain fission chains for either power production or minor actinide transmutation. Source intensity requirements span $10^{14}$--$10^{17}$~n/s depending on reactor power level, subcriticality margin, and fuel composition~\cite{Bowman_1992, Abderrahim_2010}. Current ADS neutron sources occupy two distinct regimes: compact systems delivering $10^{13}$--$10^{14}$~n/s (CPHS: 13~MeV protons on beryllium, $\sim 1 \times 10^{13}$~n/s~\cite{Wang2014_CPHS}; KIPT: 100~kW electron linac on uranium, $\sim 3 \times 10^{14}$~n/s~\cite{Gohar_OSTI, Mytsykov2019}), and large spallation facilities approaching $10^{17}$~n/s (MYRRHA: 600~MeV protons on lead-bismuth eutectic, $\sim 2 \times 10^{17}$~n/s~\cite{Abderrahim_2010, DeBruyn2021MYRRHA}).

SYNERGY occupies a previously inaccessible intermediate regime. A single uranium-target beamline is able to produces $1.3 \times 10^{15}$~n/s, an order of magnitude beyond conventional compact sources, while a 50-beamline facility reaches $6.5 \times 10^{16}$~n/s, approaching MYRRHA-class performance through \textit{distributed architecture} rather than extreme single-beam intensities. This capability enables coupling multiple independent subcritical assemblies to a single synchrotron facility, supporting parallel experimental campaigns on different fuel cycles or transmutation strategies. The fast-dominated spectrum (92\% above 100~keV) from uranium targets closely matches ADS operational requirements without spectral hardening infrastructure, while the continuous-wave beam delivery eliminates pulsed reactivity transients that complicate accelerator-reactor coupling in conventional systems.

\paragraph{Radiation Hardness Assurance}
Single event effects (SEE) in microelectronics—transient disruptions or permanent damage caused by ionizing radiation—pose critical reliability challenges for avionics, space systems, and ground-level computing infrastructure. High-energy cosmic rays interacting with Earth's atmosphere generate secondary neutron cascades with energies extending to hundreds of MeV. While the terrestrial neutron spectrum spans thermal to fast energies, MeV-range neutrons penetrate device packaging and induce nuclear recoils in silicon, producing localized ionization that flips memory states or corrupts logic circuits~\cite{Kiyanagi2021}. Accelerated radiation testing using controlled neutron sources enables reliability qualification without multi-year natural exposure, as codified in the JEDEC JESD89 standard~\cite{JESD89}.

\noindent Current testing infrastructure spans two regimes. Large spallation facilities such as ChipIr (UK) employ 700~MeV protons on tungsten to generate atmospheric-like neutron spectra optimized for SEE testing~\cite{Cazzaniga2018}, providing high-fidelity environmental simulation but limited beamtime accessibility. Compact alternatives include proton-driven systems (SHI-ATEX, Japan: 18~MeV cyclotron on beryllium~\cite{Kiyanagi2018, Kiyanagi_workshop_2019}) and electron-driven photoneutron sources (HUNS, Japan: 32~MeV linac on lead, $\sim 5 \times 10^{12}$~n/s total intensity~\cite{Sato2024}), offering dedicated access but substantially lower neutron production rates.

\noindent SYNERGY's tantalum-target configuration delivers $> 6\times 10^{14}$~n/s in the fast energy range (E$_n > 100$~keV), representing a huge intensity gain over HUNS. This enhanced intensity rate translates directly to reduced testing time: qualification campaigns requiring weeks at conventional compact facilities can be compressed to hours, dramatically improving throughput for semiconductor manufacturers. The $>$90\% fast-neutron fraction naturally matches JESD89 spectral requirements without beam shaping, while multi-beamline architecture enables simultaneous testing of multiple device families or parallel parametric studies. Furthermore, the continuous-wave beam delivery provides stable, reproducible irradiation conditions free from the beam-trip interruptions that compromise data quality in pulsed accelerator systems.

\paragraph{Radioisotope Production}
The high photon and neutron intensities achievable with SYNERGY enable dual-mode radioisotope production. The facility can operate either as a photoneutron source for $(n,\gamma)$ production routes, or exploit the primary synchrotron beam directly for $(\gamma,n)$ and $(\gamma,p)$ reactions. This approach is currently being industrialized for $^{99}$Mo/$^{99\mathrm{m}}$Tc supply in the Lighthouse project~\cite{Kramer2022} and proposed for $^{64}$Cu, $^{67}$Cu, $^{68}$Ge, and $^{186}$Re production~\cite{Schubert2024,IAEA2025radio}. For neutron-capture routes, SYNERGY's may deliver $>6 \times 10^{14}$~n/s fast neutrons per beamline with tantalum-target, or $6.0 \times 10^{14}$~n/s and $8.8 \times 10^{13}$~n/s thermal and epithermal components from heavy water targets. These intensities exceed typical compact accelerator sources ($10^{12}$--$10^{13}$~n/s)~\cite{Capogni2018,Anderson2016} while remaining below research reactors ($10^{13}$--$10^{15}$~n/(cm$^2$ s) thermal flux)~\cite{VanDerMarck2010,Konefal2022}, positioning SYNERGY as a demonstration-scale platform for $(n,\gamma)$ and $(n,2n)$ production of isotopes including $^{99}$Mo, $^{131}$I, $^{177}$Lu, and $^{186}$Re. For direct photonuclear production, the $2.02 \times 10^{17}$~$\gamma$/s photon source rate per beamline matches or exceeds electron linac bremsstrahlung systems ($10^{14}$--$10^{15}$~$\gamma$/s)~\cite{Kazakov2021,Pang2023} while providing similar spectra characteristics with respect to bremsstrahlung sources. Critically, this dual-modality capability of simultaneously accessing to both $(n,\gamma)$ and $(\gamma,n/p)$ pathways within a single infrastructure, distinguishes SYNERGY from dedicated reactor or accelerator facilities. Multi-beamline architecture further enables parallel production of different isotopes optimized to distinct reaction channels, addressing supply chain vulnerabilities through diversified production capacity rather than dependence on single-mode technologies.

\paragraph{Boron Neutron Capture Therapy}
BNCT requires epithermal neutron beams (0.5~eV--10~keV) that thermalize within tissue to maximize $^{10}$B(n,$\alpha$)$^{7}$Li reaction rates in tumor cells while sparing surrounding tissue~\cite{bookBNCT}. Clinical deployment has transitioned from research reactors, which provided suitable fluxes but severely limited accessibility, to accelerator-based neutron sources (ABNS) using low-to-medium energy proton beams (2--30~MeV) on lithium or beryllium targets. Systems such as C-BENS (Japan) and hospital-based cyclotrons (Finland) now deliver epithermal fluxes meeting IAEA clinical recommendations~\cite{Jin2022A,Bae2022Advances,Miyatake2020Boron}. Electron linac concepts remain at the feasibility stage, facing challenges in beam stability and spectral optimization~\cite{Kasilov2021Concept}, while synchrotron-driven BNCT sources are entirely absent from current literature~\cite{Verdera2024Study,Dymova2020Boron}.

\noindent SYNERGY addresses this gap through intrinsic spectral matching. Beryllium targets naturally generate epithermal-dominated spectra through in-target moderation, producing $2.3 \times 10^{14}$~n/s in the 1~eV--0.1~MeV range (Table~\ref{tab:photoneutron_flux_by_range}) without requiring external beam-shaping assemblies. Heavy water targets deliver even higher epithermal components ($8.8 \times 10^{13}$~n/s) alongside substantial thermal flux ($6.0 \times 10^{14}$~n/s), offering flexibility for different treatment protocols. The continuous-wave beam delivery eliminates the pulsed dose structure characteristic of linac systems, potentially improving treatment uniformity and reducing normal tissue toxicity. Multi-beamline architecture further enables simultaneous operation of multiple treatment rooms, addressing the patient throughput limitations of single-beam ABNS facilities. While full clinical validation—including beam-shaping design, dosimetric characterization, and radiobiological assessment—requires dedicated follow-up studies currently in preparation, these results establish synchrotron-driven photoneutron sources as viable BNCT candidates, offering institutional-scale treatment capacity through a technology pathway orthogonal to conventional proton-driven approaches.

\newcolumntype{L}{>{\RaggedRight\arraybackslash}p{2.1cm}} 
\newcolumntype{X}{>{\RaggedRight\arraybackslash\hsize=1.1\hsize}X} 
\newcolumntype{S}{>{\RaggedRight\arraybackslash\hsize=0.6\hsize}X} 

\begin{table*}[t]
\caption{Performance comparison across application domains. Comparison of representative neutron and photon facilities against achievable rates for key applications. Single beam-line SYNERGY values are derived from the present optimization study. Facility benchmarks include state-of-the-art spallation sources, research reactors, and compact accelerator-driven sources (CANS).}
\label{tab:synergy_facility_comparison}
\footnotesize 
\renewcommand{\arraystretch}{1.2}

\begin{tabularx}{\textwidth}{@{} L X X X X @{}}
\toprule
\textbf{Application} & \textbf{Spallation-Reactor} & \textbf{CANS (Proton)} & \textbf{CANS (Electron)} & \textbf{SYNERGY (this work)} \\
\midrule

\textbf{ADS} & 
\textbf{MYRRHA} (Belgium) \par
Target: LBE (Spallation) \par
$S_n \approx 2 \times 10^{17}$~n/s \par
2.4 MW (600 MeV) \cite{Abderrahim_2010} & 

\textbf{CPHS} (China) \par
Target: $^9$Be \par
$S_n \approx 10^{13}$~n/s \par
16 kW (13 MeV) \cite{Wang2014_CPHS} & 

\textbf{KIPT} (Ukraine) \par
Target: U \par
$S_n \approx 3.0 \times 10^{14}$~n/s \par
100 kW (100 MeV) \cite{Gohar_OSTI} & 

\textbf{Target:} U$_{\text{nat}}$ \par
$S_n \approx 1.3 \times 10^{15}$~n/s \par
\textit{Fast spectrum} \\
\addlinespace

\textbf{Soft-Error} & 
\textbf{ChipIr} (UK) \par
Target: W (Spallation) \par
$\phi_n \approx 4.9 \times 10^6$~n/(cm$^2$s) \par
28 kW (700 MeV) \cite{Cazzaniga2018} & 

\textbf{SHI-ATEX} (Japan) \par
Target: Be (Cyclotron) \par
$\phi_n \approx 2 \times 10^5$~n/(cm$^2$s) \par
0.36 kW (18 MeV) \cite{Kiyanagi2018} & 

\textbf{HUNS} (Japan) \par
Target: Pb \par
$S_n \approx 5.0 \times 10^{12}$~n/s \par
3.2 kW (32 MeV) \cite{Sato2024} & 

\textbf{Target:} U$_{\text{nat}}$ \par
$S_n \approx 1.2 \times 10^{15}$~n/s \par
\textit{Fast spectrum} \\
\addlinespace

\textbf{Isotopes} \par $(n,\gamma), (n,p)$ & 
\textbf{BR-2} (Belgium) \par
Target: Reactor \par
$\phi_n \approx 10^{15}$~n/(cm$^2$s) \par
Ref. \cite{hasan2020molybdenum} & 

\textbf{CPHS} (China) \par
Target: $^9$Be \par
$S_n \approx 10^{13}$~n/s \par
16 kW (13 MeV) \cite{Wang2014_CPHS} & 

\textbf{EBC} (India) \par
Target: Ta \par
$\phi_n \approx 10^7$~n/(cm$^2$s) \par
1--3 MW (10 MeV) \cite{Deo2024Measurement} & 

\textbf{Target:} U$_{\text{nat}}$ \par
$S_n \approx 1.3 \times 10^{15}$~n/s \par
\textit{High Flux} \\
\addlinespace

\textbf{Isotopes} \par $(\gamma,n), (\gamma,p)$ & 
-- & 
-- & 
\textbf{Lighthouse} (Germany) \par
Target: None \par
130 kW (75 MeV) \cite{Kramer2022} & 

\textbf{No Target} \par
$S_\gamma \approx 2.02 \times 10^{17}$~$\gamma$/s \par
\textit{Direct Beam} \\
\addlinespace

\textbf{BNCT} & 
\textbf{THOR} (Taiwan) \par
Target: Reactor \par
$\phi_n \approx 1.7 \times 10^9$~n/(cm$^2$s) \par
Ref. \cite{tung2004characteristics} & 

\textbf{NUANS} (Japan) \par
Target: Li \par
$\phi_n \approx 10^9$~n/(cm$^2$s) \par
42 kW (2 MeV) \cite{Anderson2016} & 

\textbf{PNS Project} (Iran) \par
Target: W \par
$\phi_n \approx 5 \times 10^7$~n/(cm$^2$s) \par
460 kW (20 MeV) \cite{PAZIRANDEH2011749} & 

\textbf{Target:} $^9$Be \par
$S_n \approx 2.7 \times 10^{14}$~n/s \par
\textit{Epithermal} \\

\bottomrule
\end{tabularx}
\end{table*}

%% file: Sections/05_conclusions.tex
\section*{Concluding Remarks}
We have demonstrated that synchrotron-driven photonuclear neutron production overcomes the fundamental beam power limitations constraining compact accelerator sources through architectural decoupling of charged-particle acceleration from neutron generation. By exploiting high-energy synchrotron radiation as the neutron production driver, SYNERGY achieves single-beamline intensities of $3$--$13 \times 10^{14}$~n/s and facility-scale outputs exceeding $6 \times 10^{16}$~n/s, performance previously accessible only through large spallation facilities or research reactors. The multi-beamline architecture intrinsic to synchrotron facilities extends this flexibility to simultaneous parallel operation of application-specific neutron stations within a single infrastructure, a capability inaccessible to conventional single-beam drivers.

Comprehensive Monte Carlo analysis across three independent codes (OpenMC, MCNPX, FLUKA) establishes robust performance predictions for optimized target geometries spanning low-$Z$ moderating materials (Be, BeO, D$_2$O) and high-$Z$ converters (W, Ta, U). Critically, target material selection provides inherent spectral tunability: high-$Z$ targets deliver fast-dominated spectra (92\% above 100~keV) suited to subcritical reactor coupling and radiation hardness testing, while low-$Z$ targets produce epithermal-to-thermal distributions (46--81\% below 100~keV) optimized for BNCT and neutron scattering—all without external moderation infrastructure. 

These results position synchrotron-driven sources as a credible technology pathway in the medium-to-high neutron intensity regime ($10^{14}$--$10^{16}$~n/s), bridging the gap between compact laboratory sources and centralized spallation facilities. The architectural approach offers a complementary strategy to current CANS development trends: rather than pushing single-beam systems toward increasingly challenging power densities, SYNERGY adopts mature large-scale accelerator technology to circumvent intrinsic limitations through distributed production. Future work will address engineering implementation including thermal-mechanical target design, radiation shielding optimization, and application-specific beam extraction systems, with particular focus on subcritical reactor coupling and clinical BNCT beamline development. The convergence of high-performance neutron production with synchrotron multi-user operational paradigms suggests potential for transformative impact on neutron science accessibility and diversity of applications served by single facilities.

%% file: Sections/02_methodology.tex
\section{Methods}\label{sec:methods}
\label{sec:methodology}

We performed systematic Monte Carlo simulations to optimize target geometry and quantify neutron production performance for synchrotron-driven photonuclear sources. The assessment used three independent radiation transport codes, (i.e., OpenMC, MCNPX, and FLUKA) to ensure prediction robustness against variations in nuclear data libraries and physics implementations. Parametric sweeps over target radius and length identified optimal configurations for six candidate materials spanning low-Z moderators (Be, BeO, D$_2$O) and high-Z converters (W, Ta, U$_{\text{nat}}$). Given systematic spreads observed across codes, we adopt OpenMC results with ENDF/B-VII.1 photonuclear data as the reference baseline, providing conservative performance estimates. The following sections detail the synchrotron photon source characteristics, target configurations, simulation protocols, scoring methodology, and optimization procedure.

\subsection{Synchrotron photon source specification}
\label{sec:synchrotron_methods}

The baseline SYNERGY configuration employs a 32~GeV electron storage ring with 50 independent photon beamlines distributed around a 5~km circumference (Table~\ref{tab:synchrotron_param}). Each beamline extracts synchrotron radiation from a dedicated 15~T short dipole magnet, generating a photon beam with critical energy in the MeV range optimized for photonuclear reactions. We consider two operational modes differentiated by circulating current: a 15~mA baseline configuration and a 45~mA high-current mode. The latter achieves equivalent photon power with reduced RF cavity voltage (one-third) but requires a shorter radiating magnet length (1.7~cm versus 5~cm)~\cite{chao2023handbook}. Both configurations fall within demonstrated operational regimes for large electron storage rings; the 45~mA mode is comparable to the HERA electron ring~\cite{Willeke2021}, providing confidence in technical feasibility.

\begin{figure}[htbp]
    \centering
    \includegraphics[width=0.6\linewidth]{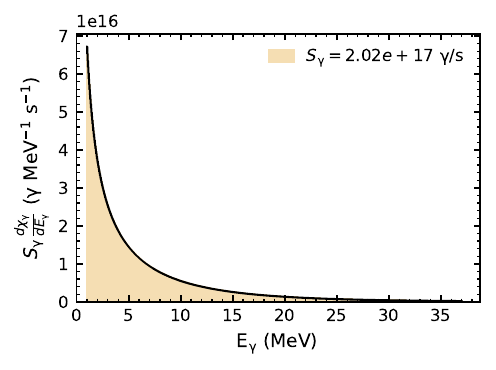}
    \caption{Energy distribution of synchrotron radiation per beamline for the 15~mA baseline configuration, normalized to 200~kW total photon beam power. The integrated photon flux in the 1--37~MeV range is $2.02 \times 10^{17}$~$\gamma$/s, encompassing the giant dipole resonance energy region for all candidate target materials.}
    \label{fig:synchrotron-flux}
\end{figure}

The photon energy distribution for each beamline was calculated using the SPECTRA code~\cite{Tanaka:wr2001} and is presented in Figure~\ref{fig:synchrotron-flux} for the baseline 15~mA configuration. The spectrum extends from 1 to 37~MeV with a total source intensity of $S_\gamma = 2.02 \times 10^{17}$~$\gamma$/s, corresponding to 200~kW photon beam power per beamline. This energy range encompasses the giant dipole resonance (GDR) peaks of target materials considered in this study (8--24~MeV for heavy nuclei, lower for light nuclei), ensuring efficient photonuclear cross-section utilization.

The choice of magnetic field strength (15~T) and electron energy (32~GeV) balances photon energy requirements against facility scale. The critical photon energy scales as $E_\gamma \propto E_e^2 \cdot B$, allowing GDR-matched photon production at manageable ring dimensions through high-field superconducting magnets rather than multi-hundred GeV electron energies. We adopt a split-solenoid design based on high-temperature superconductor (HTS) coils to achieve the 15~T dipole field while maintaining sufficient aperture for both the circulating electron beam and tangential photon extraction. The 5~km circumference supports 50 beamlines with adequate straight-section spacing ($\sim$100~m separation) for independent photon extraction and target station installation, while reducing conventional bending magnet curvature radius to minimize power consumption.

Final optimization of beam stability, quantum lifetime~\cite{chao2023handbook}, and HTS magnet engineering will be addressed in subsequent work. For the present neutron production assessment, the photon spectrum shown in Figure~\ref{fig:synchrotron-flux} serves as the input source term for all Monte Carlo simulations, with the 200~kW power level representing the normalization baseline for single-beamline performance estimates.

\begin{table}[ht]
\centering
\caption{\textbf{Baseline synchrotron parameters for SYNERGY.} Operational specifications for the 32~GeV electron storage ring. All values correspond to the 15~mA baseline configuration unless otherwise noted.}
\label{tab:synchrotron_param}
\begin{tabular}{@{}lll@{}}
\toprule
Parameter & Value & Unit \\
\midrule
Electron beam energy & 32 & GeV \\
Circulating current (baseline) & 15 & mA \\
Circulating current (high mode)\textsuperscript{a} & 45 & mA \\
Dipole magnet field & 15 & T \\
Radiating magnet length & 5\textsuperscript{b} & cm \\
Ring circumference & 5 & km \\
Number of beamlines & 50 & -- \\
Photon beam power per beamline & 200 & kW \\
Photon flux per beamline & $2.02 \times 10^{17}$ & $\gamma$/s \\
Photon energy range & 1--37 & MeV \\
\bottomrule
\end{tabular}

\smallskip
\RaggedRight \footnotesize
\textsuperscript{a} High-current mode achieves equivalent photon power with reduced RF voltage requirements.  \\
\textsuperscript{b} Reduced to 1.7~cm in the 45~mA high-current mode.
\end{table}

\subsection{Target geometry and material configurations}

\begin{figure}[t!]
    \centering
    \includegraphics[width=0.8\linewidth]{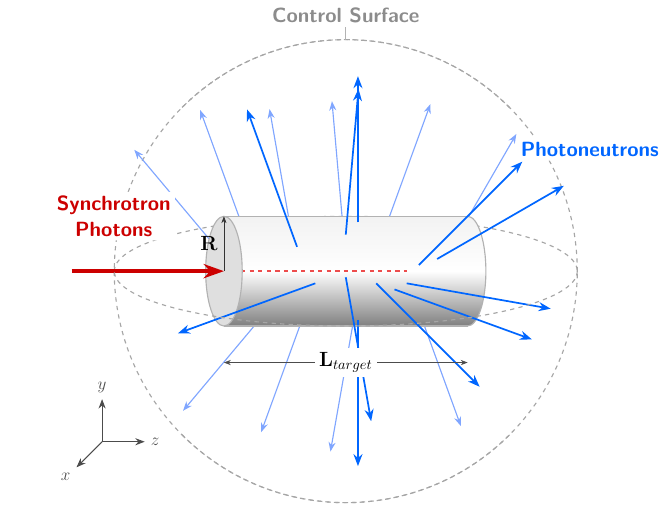} 
    \caption{Cylindrical target geometry for parametric optimization. The synchrotron photon beam impinges along the cylinder axis. Independent variation of radius ($R$) and length ($L$) identifies configurations maximizing neutron production through balanced photon attenuation and neutron transport.}
    \label{fig:target_scheme}
\end{figure}

Target geometry was parameterized as a cylinder defined by radius ($R$) and length ($L$), as illustrated in Figure~\ref{fig:target_scheme}. This two-parameter description captures competing physical processes governing neutron production efficiency: photon attenuation along the beam axis determines saturation length, while radial extent balances capture of scattered photons (from Compton scattering and electromagnetic showers) against neutron self-absorption. The incident synchrotron beam was modeled as a collimated point source aligned with the cylinder axis, isolating intrinsic material properties from beamline-specific spatial profiles.

Six candidate materials were evaluated to assess spectral versatility across application domains (Table~\ref{tab:material_composition}). Low-Z targets (Be, BeO, D$_2$O) reach their yield-optimal dimensions at thicknesses of several neutron mean free paths, so that the fast neutrons generated by the underlying photonuclear reactions undergo substantial in-target moderation before escaping, exploiting long photon mean free paths and efficient neutron down-scattering; the emerging spectrum is therefore thermalized, without requiring any external moderator. These materials target BNCT and neutron scattering applications requiring epithermal-to-thermal neutron fields. High-Z converter targets (U$_{\text{nat}}$, W$_{\text{nat}}$, $^{181}$Ta) maximize total neutron yield through $(\gamma,xn)$ reactions and photofission (for uranium); their optimal thickness is instead comparable to the photon attenuation length, insufficient to appreciably slow down the emitted neutrons, so the escaping spectrum retains its fast, evaporation-like character and is suited to subcritical reactor coupling and radiation damage studies. Natural isotopic abundances were used for all materials to reflect realistic deployment scenarios.

\begin{table}[ht]
\centering
\caption{Target material compositions and densities. Natural isotopic abundances used for all materials except monoisotopic Be and Ta. Densities correspond to 293~K, 1~atm.}
\label{tab:material_composition}
\begin{tabular}{@{}llcc@{}}
\toprule
Material & Isotope & Atomic fraction (\%) & Density (g/cm$^3$) \\
\midrule
Be & $^{9}$Be & 100 & 1.85 \\
\addlinespace
BeO & $^{9}$Be & 50 & 3.0 \\
    & $^{16}$O & 50 & \\
\addlinespace
D$_2$O & $^{2}$H  & 66.7 & 1.10 \\
      & $^{16}$O & 33.3 & \\
\addlinespace
U$_{\text{nat}}$ & $^{234}$U & 0.0054 & 19.0 \\
                & $^{235}$U & 0.72 & \\
                & $^{238}$U & 99.27 & \\
\addlinespace
W$_{\text{nat}}$ & $^{182}$W & 26.5 & 19.3 \\
                & $^{183}$W & 14.3 & \\
                & $^{184}$W & 30.6 & \\
                & $^{186}$W & 28.4 & \\
\addlinespace
$^{181}$Ta & $^{181}$Ta & 100 & 16.6 \\
\bottomrule
\end{tabular}
\end{table}

\subsection{Photonuclear reactions and nuclear data}

\begin{figure}[htbp]
    \centering
    \includegraphics[width=0.8\textwidth]{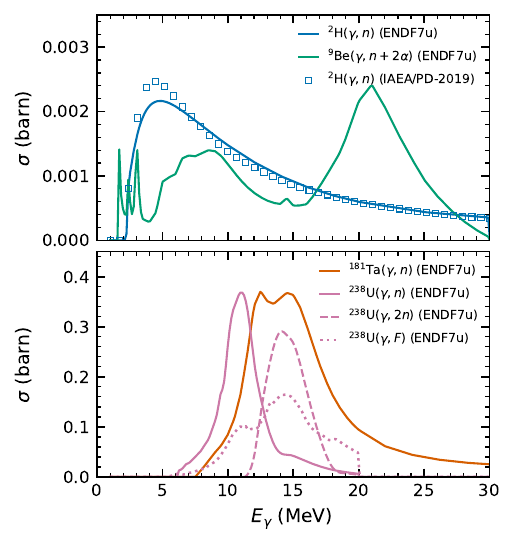} 
    \caption{\textbf{Photonuclear cross sections for low-Z and high-Z isotopes.} \textbf{Top:} Low-Z isotopes (deuterium, beryllium) with low threshold energies. Dashed line shows ENDF/B-VII.1 data; solid line shows JENDL-5 evaluation, illustrating systematic differences in deuterium cross sections~\cite{Sari2023}. \textbf{Bottom:} High-Z isotopes (tantalum, uranium) dominated by GDR peaks. For $^{238}$U, individual contributions from $(\gamma,n)$, $(\gamma,2n)$, and photofission $(\gamma,F)$ channels are shown, in good agreement with Caldwell et al. measurements~\cite{Caldwell1980}.}
    \label{fig:xs_comparison}
\end{figure}

Synchrotron-driven neutron production operates through direct photon-nucleus interactions, bypassing the electron-to-photon conversion step required in conventional electron linac systems. In the 1--37~MeV photon energy range, photonuclear reactions compete with photoatomic processes (Compton scattering, pair production) that typically dominate the total attenuation cross section. Figure~\ref{fig:xs_comparison} illustrates the photonuclear cross sections for representative low-Z and high-Z isotopes.

Low-Z materials ($^2$H, $^9$Be) exhibit low threshold energies (2.22~MeV for deuterium, 1.67~MeV for beryllium) and cross sections in the millibarn range. Compton scattering dominates competing processes, enabling photon "recycling" where scattered photons above threshold can induce subsequent reactions. High-Z materials ($^{181}$Ta, $^{238}$U) display prominent giant dipole resonance (GDR) peaks in the 10--20~MeV range with cross sections reaching hundreds of millibarns, though pair production competes strongly in this regime. Actinide converters (uranium) access additional reaction channels beyond $(\gamma,n)$: the $(\gamma,2n)$ channel and photofission $(\gamma,F)$, which releases multiple neutrons per event. For $^{238}$U, the average neutron multiplicity per fission increases approximately linearly with photon energy, reaching $\sim$2.5 at 6~MeV and $\sim$4.0 at 18~MeV due to second-chance fission above 
12~MeV~\cite{Caldwell1980,Filipescu2024}. We stress that, irrespective of target material, each individual photonuclear event releases fast, evaporation-spectrum neutrons; the thermal and epithermal components observed downstream for low-Z targets (Section~\ref{sec:results}) arise exclusively from subsequent moderation within the target volume, not from a different primary emission mechanism.

Photonuclear cross section accuracy critically impacts neutron yield predictions. Recent reviews~\cite{Sari2023} and benchmark studies~\cite{Sari2025,Garnaud2024} identify systematic discrepancies among major nuclear data libraries. The ENDF7u photonuclear library (distributed in ENDF/B-VII.1 and ENDF/B-VIII.0), while widely used in Monte Carlo codes including MCNP and OpenMC, consistently underestimates photonuclear cross sections compared to IAEA/PD-2019~\cite{IAEA_PD2019}, JENDL-5~\cite{JENDL5}, and experimental data. This bias manifests as suppressed GDR peak magnitudes and threshold mismatches, affecting both high-Z (W, Ta) and low-Z (D, Be) materials. Underestimation ranges from a few percent to factors of 2--3~\cite{Sari2023}. While uranium data show reasonable agreement with measurements, similar biases cannot be excluded.

Despite known limitations, we employ ENDF/B-VII.1 photonuclear data across all simulation codes due to its universal availability and consistent implementation. This choice introduces a systematic negative bias; reported neutron yields should be interpreted as conservative lower bounds on achievable performance. Alternative evaluations (JENDL-5, IAEA/PD-2019) would likely predict 5--10\% to factor-of-2 higher production rates depending on material and energy regime.

\subsection{Monte Carlo simulation framework}
Simulations were performed using three independent radiation transport codes to ensure prediction robustness: OpenMC v0.13.3~\cite{Romano2015}, MCNPX v2.7.0~\cite{MCNPx}, and FLUKA v4-5.0~\cite{Battistoni_2015}. This multi-code approach brackets systematic uncertainties arising from differences in physics models and transport algorithms, though OpenMC and MCNPX share common photonuclear data as detailed below.

We employed the OpenMC development branch incorporating photonuclear physics capabilities~\cite{SteinRepo}, recently verified against the broomstick benchmark problem~\cite{l_loi_2026_20803408, loi2026photonuclearneutronproductionopenmc}. Photonuclear cross sections were taken from the ENDF7u library distributed within ENDF/B-VII.1, while neutron transport employed ENDF/B-VII.0 evaluated data. The synchrotron photon spectrum (Figure~\ref{fig:synchrotron-flux}) was implemented as a fixed external source using tabulated energy distributions.

MCNPX v2.7.0 used the LA150 photonuclear library, which incorporates the same ENDF7u photonuclear evaluations as OpenMC, combined with ENDF/B-VII.0 neutron cross sections. The synchrotron spectrum was defined via the SDEF source card with energy sampling from the calculated distribution. Despite shared photonuclear data, differences in Monte Carlo transport algorithms provide partial code independence and validate the consistency of the transport implementations.

FLUKA v4-5.0 was configured to use ENDF/B-VII.1 photonuclear data where available, ensuring consistency with OpenMC and MCNPX evaluations. However, for reactions or energy regimes not covered by tabulated cross sections, FLUKA supplements evaluated data with its internal PEANUT (Precursor of Exciton Model for Nuclear Reactions) and RQMD (Relativistic Quantum Molecular Dynamics) models for photonuclear interactions. This hybrid approach, combining evaluated data with physics models, provides an independent validation pathway that is less sensitive to specific library limitations while maintaining consistency with the ENDF/B-VII.1 baseline where data exist. The synchrotron photon source was implemented through the source\_newgen.f user routine with custom energy sampling.

All codes employed analog transport without variance reduction to ensure unbiased tallies of neutron production and leakage. Statistical uncertainties were maintained below 0.2\% for all reported quantities through $10^8$ photon histories in final characterization runs.

\subsection{Neutron yield calculation and scoring}

The photoneutron yield Y$_n$ quantifies the intrinsic neutron production efficiency of a photonuclear reaction, defined as neutrons generated per incident photon. Under the assumption of single-interaction dominance (valid in the thin-target limit), Y$_n$ can be expressed analytically as:

\begin{equation}
\label{eqn:photoneutron_yield}
Y_n = \int_{0}^{\infty} \frac{d\chi_\gamma}{dE_\gamma} \left[\frac{\sigma_{\gamma,n}(E_\gamma)}{\sigma_{tot}(E_\gamma)} + 2\frac{\sigma_{\gamma,2n}(E_\gamma)}{\sigma_{tot}(E_\gamma)} + \bar{\nu}_{\gamma,F}(E_\gamma)\frac{\sigma_{\gamma,F}(E_\gamma)}{\sigma_{tot}(E_\gamma)}\right] dE_\gamma
\end{equation}

where $d\chi_\gamma/dE_\gamma$ is the incident photon spectrum, $\sigma_{tot}$ is the total photon cross section, $\sigma_{\gamma,n}$, $\sigma_{\gamma,2n}$, and $\sigma_{\gamma,F}$ are cross sections for single neutron emission, double neutron emission, and photofission, and $\bar{\nu}_{\gamma,F}$ is the average neutron multiplicity per photofission. This formulation captures the reaction probability weighted over the synchrotron spectrum but neglects neutron transport, self-absorption, and geometric effects; it is therefore only strictly valid in the optically thin limit and is not the quantity scored in the present three-dimensional Monte Carlo simulations. A detailed treatment of $Y_n$ and its dependence on nuclear data libraries is given in a companion study~\cite{loi2026photonuclearneutronproductionopenmc}.

For finite, yield-optimized target geometries, neutrons generated at depth undergo additional in-target processes, elastic and inelastic scattering, radiative capture, and, for low-Z moderators, thermalization and diffusion, before reaching the target boundary. The quantity actually scored by the Monte Carlo simulations is therefore not Y$_n$ itself but an effective, geometry-dependent yield $\tilde{Y}_n$, which folds the intrinsic photonuclear production of Eq.~\ref{eqn:photoneutron_yield} together with the net effect of in-target neutron multiplication (e.g.\ subsequent photonuclear or fission events induced by the escaping neutron flux and by scattered photons) and absorption (parasitic capture before escape). Formally,

\begin{equation}
\tilde{Y}_n \equiv \frac{1}{N_\gamma} \sum_{i=1}^{N_\gamma} \xi_i ,
\end{equation}

where $N_\gamma$ is the number of simulated photon histories and $\xi_i$ is the number of neutrons crossing the target boundary surface following photon history $i$. Surface current tallies recorded all neutrons escaping the target volume, integrating over all energies and angles. Because $\tilde{Y}_n$ is defined at the boundary of a specific, finite target geometry rather than in the thin-target limit, it is the physically appropriate and directly deployable figure of merit for engineering performance, and it is the quantity reported throughout the remainder of this work.

Spectral characterization employed logarithmic energy binning from 1~meV to 20~MeV with 50 bins. Energy-resolved tallies enabled decomposition into application-relevant groups: thermal ($\leq$1~eV), epithermal (1~eV--0.1~MeV), and fast ($\geq$0.1~MeV). As discussed above, the relative weight of these three groups in the escaping spectrum is governed by target thickness relative to the neutron mean free path, rather than by any difference in the primary emission spectrum.

The absolute neutron source rate $S_n$ (neutrons per second escaping the target) was calculated from the product:

\begin{equation}
S_n = \tilde{Y}_n \times S_\gamma
\end{equation}

where $S_\gamma = 2.02 \times 10^{17}$~$\gamma$/s is the single-beamline photon intensity. This normalization provides performance metrics directly comparable to existing neutron source technologies.

\subsection{Parametric optimization procedure}

Target optimization employed a two-stage computational strategy balancing parameter space exploration with statistical precision. First, a coarse parametric sweep over the $R$-$L$ space identified yield-maximizing regions using OpenMC with $10^7$ photon histories per configuration. Parameter ranges were material-dependent to reflect the different saturation lengths and radial extents characteristic of low-Z versus high-Z converters: high-Z materials (like tantalum) required compact dimensions with $R = 1$--$10$~cm and $L = 1$--$30$~cm due to rapid photon attenuation and neutron production saturation, while low-Z moderating materials (D$_2$O) necessitated larger volumes spanning $R = 5$--$50$~cm and $L = 5$--$300$~cm to minimize neutron leakage and maximize in-target thermalization. Intermediate materials (Be, BeO, W, U) were scanned over ranges between these extremes. All sweeps employed logarithmic spacing to efficiently sample the multi-decade dimensional domain. Statistical uncertainties in this exploratory phase remained below 1\%, sufficient to locate optimal configurations within $\pm 10\%$ geometric tolerance.

Second, once optimal dimensions $(R^{\text{opt}}, L^{\text{opt}})$ were identified for each material, high-statistics simulations with $10^8$ photon histories were performed using all three codes to precisely quantify neutron yields and spectral distributions. This two-stage workflow reduced computational cost by approximately two orders of magnitude compared to full-statistics sweeps while ensuring final performance metrics achieved sub-0.5\% statistical precision.

Cross-code agreement was evaluated through relative deviations in both integrated yield and energy-resolved spectra. For materials where codes differed by more than 10\%, the systematic spread was attributed to nuclear data variations rather than transport algorithm differences. Given that OpenMC with ENDF/B-VII.1 consistently predicted the lowest yields among the three codes, consistent with known conservative biases in this library, OpenMC results serve as the reference baseline throughout this work, providing lower-bound performance estimates with 5--10\% to factor-of-2 potential increases under alternative nuclear data evaluations.

%% file: Sections/xx_symbols.tex
\section*{Acknowledgements}

The authors would like to acknowledge Professor Enrico Padovani (Politecnico di Milano) for granting access to the MCNPX code. They also wish to thank Guy Stein for developing the OpenMC photonuclear physics branch and for his valuable support throughout the analysis. Finally, they want to thank Claudio Di Giulio (INFN), Nicola Colonna (INFN) and Giacomo Griseri (INFN e Univ. La Sapienza) for the contribution on the accelerator parameters.

\section*{Competing interests}
The authors declare no competing interests.

\section*{Data availability}
All data generated or analyzed during this study are included in this published article (and its supplementary information files).

\section*{Funding}
This research did not receive any specific grant from funding agencies in the public, commercial, or not-for-profit sectors.

\section*{Author Contribution Statement}

A.C., L.L., and A.M. conceived the project. L.L. performed the parametric optimization and primary Monte Carlo simulations with assistance from A.M. A.M. conducted the multi-code validation and independent transport simulations using the FLUKA code. D.A. and H.H.B. defined the accelerator physics parameters, including the 32~GeV storage ring configuration and the 15~T dipole magnet characteristics. L.L. handled the spectral data analysis and figure preparation. L.L. wrote the manuscript, while A.M. revised it with input from all authors. A.C. supervised the study.

\section*{List of Symbols}

\textbf{Acronyms}

\begin{longtable}{rl}
\textbf{ADS} & Accelerator Driven System \\[5pt]
\textbf{BNCT} & Boron Neutron Capture Therapy \\[5pt]
\textbf{CANS} & Compact Accelerator-driven Neutron Source \\[5pt]
\textbf{CANS-p} & Proton-driven Compact Accelerator-driven Neutron Source \\[5pt]
\textbf{CANS-e} & Electron-driven Compact Accelerator-driven Neutron Source \\[5pt]
\textbf{GDR} & Giant Dipole Resonance \\[5pt]
\textbf{LINAC} & Linear Accelerator \\[5pt]
\textbf{PEDD} & Peak Energy Deposition Density \\[5pt]
\textbf{RF} & Radio Frequency \\[5pt]
\textbf{SEE} & Single Event Effects \\[5pt]
\textbf{SYNERGY} & Synchrotron-driven NEutron source for Research, enerGY and therapY
\end{longtable}
\addtocounter{table}{-1}

\textbf{Greek Symbols}

\begin{longtable}{rl}
{$\gamma$} & Photon (gamma radiation) \\[5pt]
{$\sigma$} & Microscopic cross section \\[5pt]
{$\sigma_{\text{tot}}$} & Total microscopic photon interaction cross section \\[5pt]
{$\sigma_{\gamma,n}$} & Photonuclear single-neutron emission cross section \\[5pt]
{$\sigma_{\gamma,2n}$} & Photonuclear double-neutron emission cross section \\[5pt]
{$\sigma_{\gamma,f}$} & Photonuclear fission cross section \\[5pt]
{$\nu_{\gamma,f}$} & Average number of neutrons emitted per photofission \\[5pt]
{$\phi_n$} & Neutron flux
\end{longtable}
\addtocounter{table}{-1}

\textbf{Latin Symbols}

\begin{longtable}{rl}
{$B$} & Magnetic field strength at photon source point \\[5pt]
{Y$_n$} & Photoneutron yield (neutrons per source photon) \\[5pt]
{S$_n$} & Photoneutron source rate (neutrons per second) \\[5pt]
{S$_\gamma$} & Incident photon source strength \\[5pt]
{$N_\gamma$} & Number of simulated source photons \\[5pt]
{$E_\gamma$} & Photon energy \\[5pt]
{$E_e$} & Electron energy \\[5pt]
{$R$} & Target radius \\[5pt]
{$L$} & Target length \\[5pt]
{$P$} & Beam or photon power \\[5pt]
{$A$} & Mass number \\[5pt]
{$Z$} & Atomic number
\end{longtable}
\addtocounter{table}{-1}